\documentstyle[aps,12pt]{revtex}

\begin{document}
\author{O.B.Zaslavskii}
\address{Department of Physics, Kharkov State University, Svobody Sq.4, Kharkov\\
310077, Ukraine\\
e-mail: olegzasl@aptm.kharkov.ua}
\title{Horizon/matter systems near the extreme state }
\maketitle

\begin{abstract}
It is shown that in the extreme limit with a zero surface gravity but
nonzero local temperature the limiting metric of a generic static black hole
is determined by a metric induced on a horizon and one function of two
coordinates, stress-energy tensor of a source picking up its values from a
horizon. The limiting procedure is extended to rotating black holes. If the
extreme limit is due to merging a black hole horizon and cosmological one
both horizons are always in thermal equilibrium in this limit. This is
proved for a generic case of static or axially-symmetrical rotating
spacetimes.
\end{abstract}

\draft


\section{introduction}

The significant part of physically relevant black hole and cosmological
spacetimes (such as the Reissner-Nordstr\"{o}m, Schwarzschild-de Sitter,
Kerr metrics) possesses an extreme state. Either geometrical or
thermodynamic properties of such a state differ from nonextreme one in an
essential way: a surface gravity and a corresponding Hawking temperature are
equal to zero, the proper distance from a horizon to any other point
diverges, the entropy assigned to a horizon is zero \cite{ross}. Thus, the
properties of the extreme state are, according to the prescription made in 
\cite{ross}, highly nontrivial. On the other hand, the limiting transition
from the topological sector of nonextreme black holes to the nonextreme one
seemed, for a first glance, to be obvious. As, by definition, the surface
gravity goes to zero as one approaches the state in question, one could
expect that the limiting state is simply a state with the zero temperature
and without profound thermodynamic properties and, apart from this, the
proper distance between a horizon and any other point tends to infinity in
accordance with known properties of the extreme black hole (the simplest
example is the extreme Reissner-Nordstr\"{o}m black hole). It turned out,
however, that this limit can be performed in a rather nontrivial way, so the
temperature and the above mentioned proper distance both remain finite \cite
{zasl96}, \cite{zasl97}, \cite{zaslrad}.

The main point here can be explained from the thermodynamic viewpoint and
consists in the crucial difference between the temperature $T_{0}$ measured
at infinity (its role in black-hole physics is played by the Hawking
temperature $T_{H}$) and Tolman temperature $T$ in a static spacetime
measured by a local observer. In general relativity they are connected by
the relationship $T=T_{0}/\sqrt{-g_{00}}$. Here $T$ is a local (Tolman)
temperature, $T_{0\text{ }}$is that measured at infinity. For nonextreme
black hole $T_{0\text{ }}$coincides with the Hawking temperature $T_{H}$. In
the extreme limit $T_{H}\rightarrow 0$ by definition, so it would seem
obvious that this limit is trivial in the sense that $T\rightarrow 0$
everywhere. Meanwhile, it was shown in \cite{zasl96} that the limit under
discussion can be taken in such a way that simultaneously with $T_{H\text{ }}
$the quantity $g_{00\text{ }}$goes to zero as well, the local temperature $T$
remaining nonzero. It is worth stressing that both temperatures have
qualitatively different status in gravitational thermodynamics: it is the
quantity $T$ which determines the properties of the canonical or grand
canonical ensemble, thus having direct physical meaning whereas $T_{0}$ in
the framework of these ensembles can be rescaled in arbitrary way \cite
{york86}. Therefore, it is essential that the limiting state is achieved at
a finite $T$. As was demonstrated in \cite{zasl96} for the
Reissner-Nordstr\"{o}m metric, the radial coordinate in this limit becomes
degenerate, so all points pick up the value of it equal to that of the
horizon. If, however, this coordinate is properly rescaled (along with a
time coordinate) the resulting geometry is well-defined and has an universal
form \cite{zasl97}. It turns out also that the limiting state possesses
further interesting properties: although in the limit under discussion
temperature $T\neq 0$, the entropy of Hawking radiation surrounding a
horizon equals zero \cite{zaslrad}, quantum corrections to the entropy of a
black hole itself have the universal form \cite{mannsol}. 

In the present paper we generalize the treatment of the extreme limit in
three directions. First, we find the limiting form of a metric of a generic
static distorted nonextreme black hole near the extreme state. Second, we
include in consideration systems which possess simultaneously a black hole
and cosmological horizons and show that in the limit when metrics induced on
both horizons coincide (in particular, for spherically-symmetrical systems
radii of both horizons merge) the state of thermal equilibrium arises
irrespectively of details of a metric and matter distribution between
horizons. Third, we apply our approach to rotating horizons and find the
limiting metric near the state under discussion. Apart from general
motivation connected with the role of the extreme state in black hole
physics, the issue addressed in the present paper establishes relationship
between different types of metrics in general relativity which are obtained
one from another by changing parameters of solutions and, in fact, realizes
the case when finding ''limits of spacetimes'' \cite{geroch}, \cite{paiva}
is motivated physically from the viewpoint of black hole thermodynamics.

\section{extreme limit for a generic static black hole}

Consider a generic static black hole without any assumptions about a
spherical (or any other) symmetry. As in a general case there is no analog
of radial coordinate $r$ we will use another approach. Let us choose the
coordinate system which consists of equipotential surfaces $g_{00}=const$
and coordinates on each surface. (Such a system was used while proving the
uniqueness theorem for static black holes in an asymptotically flat
spacetimes \cite{israel}. The latter assumption is not relevant for us.)
Then a metric can be written in the form 
\begin{equation}
ds^{2}=-dt^{2}V^{2}+\chi ^{-2}dV^{2}+\gamma _{ab}dy^{a}dy^{b}
\label{distorted}
\end{equation}
Here $a,b=1,2;$ components $\gamma _{ab}$ and $\chi $ depend, generally
speaking, on all three spacelike coordinates. The quantity $\chi
_{0}=\lim_{V\rightarrow 0}\chi $ is a constant \cite{israel} equal to the
surface gravity \cite{carter} which determines the Hawking temperature
according to $T_{H}=(2\pi )^{-1}\chi _{0}.$

As before, we will consider such a limiting transition in the process of
which a new manifold is obtained from the vicinity of a horizon. Similarly
to \cite{zasl96}, \cite{zasl97}, we demand that a local Tolman temperature $%
T=T_{H}/\sqrt{-g_{00}}$ remain finite in each point outside a horizon.
Therefore, let simultaneously $V\rightarrow 0$ and $\chi _{0}\rightarrow 0$
in such a way that the ratio $V/\chi _{0}$ be finite. It is convenient to
introduce new coordinates according to 
\begin{equation}
x=V/\chi _{0},\text{ }t_{1}=t\chi _{0}  \label{distcoord}
\end{equation}
For small $\chi _{0}$ we can use the following expansion for $\chi $: 
\begin{equation}
\chi ^{2}(y^{a},V)=\chi ^{2}(y^{a},x\chi _{0})=\chi _{0}^{2}+(x\chi
_{0})^{2}f(y^{1},y^{2})+....  \label{exp}
\end{equation}
In the limit under consideration linear terms in (\ref{exp}) are missing
since they would have destroyed the regularity of a metric when $\chi
_{0}\rightarrow 0$ as follows from substitution of (\ref{exp}) into (\ref
{distorted}) (cf. \cite{and} where power expansion for metric coefficients
was used for the analysis of the regularity of spherically-symmetrical
metrics).

As a result, we find the limiting form of a metric for $\chi _{0}\rightarrow
0$: 
\begin{equation}
ds^{2}=-dt_{1}^{2}x^{2}+\frac{dx^{2}}{1+x^{2}f(y^{1},y^{2})}+\gamma
_{ab}^{h}dy^{a}dy^{b}  \label{distlim}
\end{equation}

It is supposed here that the function $f>0,$ $\gamma
_{ab}^{h}(y^{1},y^{2})=\gamma _{ab}(y^{1},y^{2},V=0).$ In so doing, a metric
may be vacuum or have nonzero stress-energy tensor as a source. In the
latter case the values of its components in the orthogonal frame can be
obtained from the corresponding values for an original metric (\ref
{distorted}) by putting $V=0.$ In other words, this tensor picks up its
values from a horizon of the metric (\ref{distorted}).

It may happen that, depending on properties of the function $f,$ the
limiting metric may acquire elements of symmetry which are absent in an
original metric (\ref{distorted}). For instance, if a two-dimensional
surface $x=const$ represents a sphere and $f=r_{+}^{-2}$ the metric (\ref
{distlim}) by virtue of the substitution can be cast into the form 
\begin{equation}
ds^{2}=r_{+}^{2}(-dt_{1}^{2}\sinh ^{2}x+dx^{2}+d\omega ^{2})
\label{spherlim}
\end{equation}
If $f=0$ the metric in the $(x,t_{1})$ sector is the Rindler spacetime.
These examples agree with results of \cite{zasl97}. In both cases the metric
represents a direct product of two-dimensional spaces. The latter property
holds true for any (not necessarily spherically-symmetrical) metric with $%
f=const.$

Thus, we see that properties of limiting metrics are determined in fact by a
single function $f$, a metric induced on an arbitrary surface $x=const$
being determined by the metric on a horizon of an original spacetime. As a
rule, the metric of a four-dimensional black hole distorted by matter or
external sources cannot be found exactly. However, formulae and reasonings
listed above show that even though an original metric is not known one
manages to obtained the limiting form of it near the extreme state. In other
words, approaching the extreme state restricts the variety of metrics
according to (\ref{distlim}). To conclude this section, one reservation is
in order. It was assumed above that a metric on a horizon is regular.
Meanwhile, for extreme dilatonic black holes the horizon surface tends to
zero \cite{dilaton}, so this case is not encompassed by the approach
developed above. Nevertheless, the limiting transition to a regular metric
in the topological sector of nonextreme black holes in this case does exist 
\cite{zasldil}.

\section{distorted black-hole and cosmological horizons in thermal
equilibrium}

It was assumed in the preceding section that a system possesses only a black
hole horizon. The extreme limit implying that $T_{H}\rightarrow 0$ could be
realized due to merging a black hole and inner horizons as it takes place,
for instance, in the Reissner-Nordstr\"{o}m metric. Meanwhile, there exists
also another class of examples of physical interest in which, along with a
black hole horizon, a cosmological one is present and the extreme limit is
achieved due to merging both horizons. The interesting point here consists
in that, although radii of horizons coincide, the proper distance between
them remain finite, the Hawking temperatures associated with both horizons
being equal, so a black hole and cosmological horizons attain the thermal
equilibrium in the extreme limit. The corresponding observation were made,
among static metrics, for a spherically-symmetrical spacetimes only and,
moreover, only for the several particular classes of them - such as the
Schwarzschild - de Sitter and Reissner-Nordstr\"{o}m - de Sitter metrics 
\cite{perry}, \cite{moss}, \cite{roman}, \cite{mann} (plus the Kerr- de
Sitter metric for the case with rotation \cite{page} - see next section).
The aim of the present section is to generalize the corresponding results to
generic static black hole - cosmological systems with distorted horizons.

We will use the approach similar to that of the preceding section. Consider
the metric 
\begin{equation}
ds^{2}=-dt^{2}f(V)+\chi ^{-2}dV^{2}+\gamma _{ab}dy^{a}dy^{b}  \label{2hor}
\end{equation}
with $f(V)=V^{2}(V_{0}^{2}-V^{2})g(V)$ where the function $g$ is regular and
nonzero near a black hole horizon $V=0$ and a cosmological one $V=V_{0}$.
For the metric to be regular near $V=V_{0}$, the function $\chi $ must have
the form $\chi ^{2}=(V_{0}^{2}-V^{2})h(V)$ where $h$ is regular near $V_{0}.$
In general, the Hawking temperatures associated with both horizon are
different, so the thermal equilibrium is impossible.

Consider now the limit $V_{0}\rightarrow 0$ in which both horizons seemingly
coincide. Introducing the new variables according to $V=V_{0}z$ and $%
t=t_{1}V_{0}^{-2}[h(0)g(0)]^{-1/2}$ we obtain 
\begin{equation}
ds^{2}=h(0)^{-1}[-dt_{1}^{2}z^{2}(1-z^{2})+\frac{dz^{2}}{1-z^{2}}]+\gamma
_{ab}^{h}dy^{a}dy^{b}  \label{limit2}
\end{equation}
where $\gamma _{ab}^{h}=\lim_{V\rightarrow 0}\gamma _{ab}$. By substitution $%
z=\sin \frac{x}{2}$ this metric can be cast to the form 
\begin{equation}
ds^{2}=[4h(0)]^{-1}[-dt_{1}^{2}\sin ^{2}x+dx^{2}]+\gamma
_{ab}^{h}dy^{a}dy^{b}  \label{2limit2}
\end{equation}
It is easily seen from (\ref{2limit2}) that the horizons are located at $x=0$
and $x=\pi $, so the proper distance between them remains finite. For the
Euclidean version of this metric the choice of the period equal to $2\pi $
makes the metric regular simultaneously near both horizons, so both horizon
have the same temperature and are in the state of thermal equilibrium.

It is worth stressing two crucial points. First, the conclusion derived
above is valid irrespectively of the presence of matter between both
horizons and the particular properties of its stress-energy tensor. It
means, in particular, that the backreaction of quantum field on the
Schwarzschild - de Sitter or Reissner-Nordstr\"{o}m - de Sitter metric does
not destroy the state of thermal equilibrium between both horizons in the
extreme limit. In so doing, this limit is to be understood as the
coalescence of radii $r_{+}$ and $r_{-}$ where these quantities themselves
are related to the system ''dressed'' by Hawking radiation. Second, the
general structure (\ref{2limit2}) follows from the regularity of the
manifold and does not use field equations, so it is model-independent. Thus,
self-consistent quantum-corrected geometries do exist in the extreme limit
under discussion. Whereas their general form is determined by eq.(\ref
{2limit2}) the other details such as the coefficient $h(0)$ cannot be found
without taking into account field equations. Carrying out backreaction
program in such a situation is beyond the scope of the present paper and
deserves separate treatment. Here we only mention that the self-consistent
solutions of field equations similar to the $(x$, $t_{1})$ part of the (\ref
{2limit2}) exist in two-dimensional dilaton gravity \cite{zasl98}. The
existence and properties of self-consistent solutions with backreaction of
quantum radiation can be, in particular, of interest from the viewpoint of
investigating dynamics of (anti-)evaporation of Schwarzschild-de Sitter
black holes \cite{bousso}. The obtained form of the limiting geometry may be
also of interest for the effect of production of black hole pairs in
cosmology \cite{mann}.

Each of the horizons contributes the term $A/4$ ($A$ is the surface area of
the horizon) into the entropy which is equal to their sum, both
contributions being equal. It is instructive to note here the difference
with the situation when the event horizon and the inner one merge whereas
the event horizon and the cosmological one do not. Let for simplicity the
system is a spherically-symmetrical and $r_{-}$, $r_{+\text{ }}$and $r_{c}$
correspond to the inner, event and cosmological horizon, respectively. Let $%
r_{-}\rightarrow r_{+}\neq r_{c}$. As is shown in \cite{zasl96}, \cite
{zasl97}, the limiting metric has only one horizon. As an original metric
(with arbitrary $r_{-}<r_{+}<r_{c}$) had either the black hole or
cosmological horizon one may wonder where the entropy associated with a
cosmological horizon got to. The point, however, is that all points of the
Euclidean manifold have $r=r_{+}$, so the cosmological horizon as well as
any points with $r>r_{+}$ do not belong to the manifold at all. As a result,
only the horizon $r=r_{+}$ contributes to the entropy. The zero-loop entropy
of the original spacetime associated with horizons is $S_{0}=$ $\pi
(r_{+}^{2}+r_{c}^{2})$. The entropy of the final state after the limiting
transition is performed is not $S_{0}$ with $r_{+}$ and $r_{c\text{ }}$%
replaced by their limiting values, but is equal to $\pi r_{+}^{2}$ instead.
On the other hand, for the case described by eq.(\ref{2limit2}) $r_{+}=r_{c}$
and the entropy does equal $\pi (r_{+}^{2}+r_{-}^{2})=2\pi r_{+}^{2}$.

\section{extreme limit for nonextreme rotating black holes}

Discuss now the case of four-dimensional rotating black holes. It is easy to
carry out the corresponding limiting procedure to the extreme state for an
axially-metric of a general form: 
\begin{equation}
ds^{2}=-dt^{2}f+(N^{\phi }dt+d\phi )^{2}g_{\phi \phi
}+g_{rr}dr^{2}+g_{\theta \theta }d\theta ^{2}  \label{rot}
\end{equation}
where all metric coefficients depend on $r$ and $\theta $ only. Let they
have the form 
\begin{equation}
f=(r-r_{+})(r-r_{-})\mu (r,\theta )\text{, }g_{rr}=[(r-r_{+})(r-r_{-})%
\lambda (r,\theta )]^{-1}\text{, }N^{\phi }=(r-r_{+})\eta (r,\theta )
\label{form}
\end{equation}
where $r_{+}>r_{-\text{ }}$ corresponds to the event horizon and functions $%
\mu $, $\lambda $, $\eta $ are finite near $r_{+}.$ In particular, all
properties indicated above are inherent to the Kerr and Kerr-Newman metrics
in the frame rotating with respect to a distant observer with the angular
velocity equal to that of a black hole. The Hawking temperature for a black
hole spacetime described by the metric (\ref{rot}) is equal to $T_{H}=(4\pi
)^{-1}(r_{+}-r_{-})\alpha $ where $\alpha =\sqrt{\lambda ^{h}\mu ^{h}\text{,}%
}$ the index ''h'' indicates that the corresponding quantities are to be
taken at the horizon. The quantity $\alpha $ is constant on the horizon
surface due to the constancy of a surface gravity.

Consider what is happening in the limit $r_{+}\rightarrow r_{-}$. Let us
make the substitution 
\begin{equation}
r-r_{+}=(r_{+}-r_{-})\sinh ^{2}\frac{x}{2}\text{, }t=2t_{1}[(r_{+}-r_{-})%
\alpha ]^{-1}  \label{subst}
\end{equation}
Then the metric in the limit at hand reads 
\begin{equation}
ds^{2}=\lambda _{h}^{-1}(-dt_{1}^{2}\sinh ^{2}x+dx^{2})+g_{\phi \phi
}^{h}(d\phi +dt_{1}\sinh ^{2}\frac{x}{2}\eta ^{h})^{2}+g_{\theta \theta
}^{h}d\theta ^{2}  \label{limrot}
\end{equation}
In particular, substituting into (\ref{rot}), (\ref{limrot}) the explicit
values of coefficients for the Kerr metric we obtain 
\begin{equation}
ds^{2}/a^{2}=(1+\cos ^{2}\theta )(-dt_{1}^{2}\sinh ^{2}x+dx^{2}+d\theta
^{2})+4\frac{\sin ^{2}\theta }{(1+\cos ^{2}\theta )}(d\phi +2\sinh
^{2}x/2dt_{1})^{2}  \label{kerr}
\end{equation}
where $a$ is the angular momentum parameter of the Kerr black hole. It is
nothing else than the rotating analog of the Bertotti-Robinson metric \cite
{bh}. In a similar manner for the extreme limit of the nonextreme
Kerr-Newman black holes the metric is 
\begin{equation}
ds^{2}=(r_{+}^{2}+a^{2}\cos ^{2}\theta )(-dt_{1}^{2}\sinh
^{2}x+dx^{2}+d\theta ^{2})+\frac{(r_{+}^{2}+a^{2})^{2}}{r_{+}^{2}+a^{2}\cos
^{2}\theta }\sin ^{2}\theta [d\phi +4(\frac{r_{+}a}{r_{+}^{2}+a^{2}})\sinh
^{2}\frac{x}{2}dt_{1}]^{2}  \label{newman}
\end{equation}

For uncharged extreme black holes $r_{+}=a$ and (\ref{newman}) turns into (%
\ref{kerr}). For static holes $a=0$ and (\ref{newman}) turns into the finite
temperature version of the Bertotti-Robinson spacetime in agreement with the
previous result \cite{zasl97}.

The electromagnetic field for the spacetime (\ref{newman}) is obtained in
the limit under discussion from that of the Kerr-Newman in a straightforward
manner. The result reads 
\begin{equation}
{\bf F}=\frac{2Qar_{+}\cos \theta \sin \theta (r_{+}^{2}+a^{2})}{%
(r_{+}^{2}+a^{2}\cos ^{2}\theta )^{2}}d\theta \wedge (d\phi +\frac{4ar_{+}}{%
r_{+}^{2}+a^{2}}\sinh ^{2}\frac{x}{2}dt)+\frac{Q\sinh x(r_{+}^{2}-a^{2}\cos
^{2}\theta )}{(r_{+}^{2}+a^{2}\cos ^{2}\theta )^{2}}dx\wedge dt_{1}
\end{equation}

The next example is a class of metric for which either effect of rotation or
a cosmological horizon is present. It means that the metric we start with
has now the form (\ref{rot}) but instead of (\ref{form}) the metric
coefficients obey the conditions 
\begin{equation}
f=(r-r_{1})(r_{2}-r)\mu \text{, }g_{rr}=(r-r_{1})(r_{2}-r)\lambda \text{, }%
N^{\phi }=(r-r_{1})\eta  \label{formrot}
\end{equation}
Then, repeating calculations step by step we arrive at 
\begin{equation}
ds^{2}=\lambda _{h}^{-1}(-dt_{1}^{2}\sin ^{2}x+dx^{2})+g_{\phi \phi
}^{h}(d\phi +\eta ^{h}\sin ^{2}\frac{x}{2}dt_{1})+g_{\theta \theta
}^{h}d\theta ^{2}  \label{2hrot}
\end{equation}
Similarly to the case of non-rotational metrics, both horizon are situated
at the finite proper distance and have (in the limit in question) equal
temperatures. In particular, the form (\ref{2hrot}) includes the extreme
limit of the Kerr-de Sitter metric obtained earlier in \cite{page}.

\section{summary and conclusion}

We have developed rather general unified approach to the treatment of the
extreme state (state with a zero surface gravity) within the nonextreme
topological sector. It is based on the physical demand that a local Tolman
temperature be finite at any point outside a horizon. The corresponding
limiting procedure is well-defined and carrying it out shows that a thin
layer adjoining an event horizon develops into a new manifold. In so doing,
a proper distance between a horizon and any other point outside it remains
finite. If a metric is non-vacuum the stress-energy tensor of the resulting
metric picks up its values from a horizon of an original metric, so it does
not depend on a distance from a horizon. The approach presented in this
paper enabled us to handle on equal footing the cases when the extreme limit
arises either due to coalescing an event and inner horizons of a black hole
or when it is due to merging of a black hole and cosmological horizons. We
have manage to relax the condition of a spherical symmetry and obtain a
general form of the limiting metric of a static black hole which is
determined by two entities. The first one is a two-dimensional metric
induced on a horizon of an original metric which plays a role of a metric
induced on an equipotential surface $g_{00}$ of the limiting metric,
component of it being the same for any such a surface. The second one is a
single function of two variables (coordinates on a surface $g_{00}=const)$
only. Thus, even without complete knowledge of a solution which possesses
the extreme limit one may gain information about the limiting form of it.
For rotating black holes the limiting metric obtained from the Kerr and
Kerr-Newman solutions is obtained explicitly and has a rather simple form.

While black holes themselves in a sense possess universality due to
uniqueness and no-hair theorems (strong gravitational field deletes
information about details of a system structure, so only a few parameters
characterize black hole spacetimes) the extreme limit turns out to be even
''more universal'' as compared to other black hole configurations. Owing
this a number of general conclusion have been made above irrespectively of
details of a system.

\end{document}